\documentclass{iopart}
\usepackage{iopams}
\usepackage{graphicx}
\begin{document}
\title{Strong lensing probability in TeVeS theory}
\author{Da-Ming Chen}
\address{National Astronomical Observatories, Chinese Academy of
Sciences, Beijing 100012, China} \ead{cdm@bao.ac.cn}
\begin{abstract}
We recalculate the strong lensing probability  as a function of the
image separation in TeVeS (tensor-vector-scalar) cosmology, which is
a relativistic version of MOND (MOdified Newtonian Dynamics). The
lens is modeled by the Hernquist profile. We assume an open
cosmology with $\Omega_b=0.04$ and $\Omega_\Lambda=0.5$ and three
different kinds of interpolating functions. Two different galaxy
stellar mass functions (GSMF) are adopted: PHJ
(Panter-Heavens-Jimenez, 2004) determined from SDSS data release one
and Fontana (Fontana et al., 2006) from GOODS-MUSIC catalog. We
compare our results with both the predicted probabilities for lenses
by Singular Isothermal Sphere (SIS) galaxy halos in LCDM (lambda
cold dark matter) with Schechter-fit velocity function, and the
observational results of the well defined combined sample of Cosmic
Lens All-Sky Survey (CLASS) and Jodrell Bank/Very Large Array
Astrometric Survey (JVAS). It turns out that the interpolating
function $\mu(x)=x/(1+x)$ combined with Fontana GSMF matches the
results from CLASS/JVAS quite well.
\end{abstract}
 \submitto{JCAP}
\maketitle
\section{Introduction}
The standard LCDM cosmology is very successful in explaining the
cosmic microwave background (CMB, see, e.g., \cite{spergel07}),
baryonic acoustic oscillation (BAO, see, e.g., \cite{eisenstein05}),
gravitational lensing (see, e.g., \cite{kochanek04}) and large scale
structure (LSS) formation. However, LCDM faces some fundamental
difficulties. From the observational point of view, the challenges
to LCDM arise from smaller scales. For example, the theory cannot
explain Tully-Fisher law and the Freeman law
\cite{dalcanton97,vandb2000}. The most difficult ones are the
satellites problem and cusps problem. The most key problems are, of
course, the unknown nature of Dark Matter (DM) and Dark Energy (DE).
Before CDM particles are detected in the lab, science should remain
open to the prospect that DM (and for the similar reasons, DE)
phenomena may have some deep underlying reason in new physics.

There are several proposals for resolving DM and DE problems by
modifying Newtonian gravity or general relativity (GR) rather than
resorting to some kinds of exotic matter or energy. MOND
\cite{milgrom83} was originally proposed to explain the observed
asymptotically flat rotation curves of galaxies without DM, however,
it was noticed that MOND can also explain Tully-Fisher law and
Freeman law \cite{mcgaugh98,mcgaugh2000}. It is believed that MOND
is successful at galactic scales \cite{wxf07,zhao07} (but see
\cite{klypin07} for satellites problem). The challenges to MOND
arise from clusters of galaxies \cite{sanders02}, in which, some
kind of dark matter, possibly some massive neutrinos with the mass
of $\sim 2$ev, is also needed to explain the dynamics of
galaxies\cite{angus07b}. MOND and its relativistic version, TeVeS
\cite{bekenstein04}, are only concerned with DM, remain DE as it is.
By adding a $f(R)$ term in Einstein-Hilbert Lagrangian, where $R$ is
the Ricci scalar, the so called $f(R)$ gravity theory can account
for DE
\cite{abfo05,cct03,cdtt04,no03,no07,sotiriou06a,sotiriou06b,vollick03}.
Another interesting theory is Modified Gravity (MOG)
\cite{moffat06}, it is a fully relativistic theory of gravitation
that is derived from a relativistic action principle involving
scalar, tensor and vector fields. MOG has been used successfully to
account for galaxy cluster masses \cite{bm06a}, the rotation curves
of galaxies (similar to MOND) \cite{bm06b}, velocity dispersions of
satellite galaxies \cite{mt07a}, globular clusters \cite{mt07b} and
Bullet Cluster \cite{bm07}, all without resorting to DM. Most
recently, MOG is used to investigate some cosmological observations
(CMB, galaxy mass power spectrum and supernova), and it is found
that MOG provides good fits to data without DM and DE \cite{mt07c}.

Any modifications to traditional gravity theory must be tested with
observational experiments. Gravitational lensing provides a powerful
probe to test gravity theory \cite{schneider92,wu96}. It is well
known that, in standard cosmology (LCDM), when galaxies are modeled
by a SIS and galaxy clusters are modeled by a Navarro-Frenk-White
(NFW) profile, the predicted strong lensing probabilities can match
the results of CLASS/JVAS quite well
\cite{chae03,chena,chenb,chenc,chend,inada03,keeton01,kochanek01,
koopmans06,li02, lgl07,lopes, mitchell05,
oguri02,oguri03a,oguri03b,oguri04a,oguri04,oguri06,peng06,rusin05,sarbu01,
wj04,zhang2004,zhang05}.

This paper is devoted to explore the strong lensing statistics in
TeVeS theory. As an alternative to LCDM cosmology, TeVeS cosmology
has received much attention in the  recent literature, in particular
in the aspect of gravitational lensing \cite{angus06,chiu,zhao06a},
for reviews see \cite{bekenstein06,sanders06}. Before TeVeS, strong
gravitational lensing in the MOND regime could only be manipulated
by extrapolating non-relativistic dynamics \cite{qin,mortlock}, in
which the deflection angle is only half the value in TeVeS
\cite{zhaoqin06}. In TeVeS theory, it is now established that, for
galaxy clusters, both weak and strong lensing need extra DM to
explain observations \cite{angus07a,feix07,famaey07,takahashi07},
possibly neutrinos with the mass of $\sim 2$ev, like the dynamics of
galaxies. The situation is better for galaxies, as will be shown in
this paper. In our previous paper \cite{chen06}, as a first try to
calculate the strong lensing probability as a function of the
image-separation $\Delta\theta$ in TeVeS cosmology, we assumed a
flat cosmology with $\Omega_b=1-\Omega_\Lambda=0.04$ and the
simplest interpolating function $\mu(x)={\rm min}(1,x)$. In this
paper, we assume an open cosmology with $\Omega_b=0.04$ and
$\Omega_\Lambda=0.5$ and three different kinds of interpolating
functions. As for mass function, in addition to the PHJ GSMF
\cite{panters} used in our previous paper, we also adopt a
redshift-dependent Fontana GSMF \cite{fontana06}. Further more, the
amplification bias is calculated based on the total magnification of
the outer two brighter images rather than the magnification of the
second bright image of the three images as did in our previous work
\cite{chen06}.

\section{TeVeS cosmology and deflection angle}
Gravitational lensing can be used to test TeVeS in two aspects.
First, in the non-relativistic and spherical limit, TeVeS reduces to
MOND. The deflection angle of the light ray passing through the
lensing object can be calculated in MONDian regime (this will be
discussed later). Second, the distances between the source, the lens
and the observer are cosmological and thus depend on the geometry
and evolution properties of the background universe. As argued by
Bekenstein \cite{bekenstein04,zhao06a}, the scalar field $\phi$,
which is used to produce a MONDian gravitational acceleration in
non-relativistic limit, contributes negligibly to Hubble expansion.
According to the cosmological principle, the physical metric takes
the usual Friedmann-Robertson-Walker (FRW) form in TeVeS
\cite{bartelmann01},
\begin{equation}
d\tau^2=-c^2\rmd
t^2+a(t)^2[\rmd\chi^2+f^2_{K}(\chi)(\rmd\theta^2+\sin^2\theta
\rmd\psi^2)],
\end{equation}
where $c$ is the speed of light, $a(t)$ is the scale factor and
\begin{equation}
f_{K}(\chi)=\cases{K^{-1/2}\sin(K^{1/2}\chi) & $(K>0)$ \cr \chi
&$(K=0)$. \cr (-K)^{-1/2}\sinh[(-K)^{1/2}\chi] & $(K<0)$}
\end{equation}
As in general relativity (GR), we define the cosmological
parameters:
\begin{equation}
\Omega_{\rm b}\equiv\frac{\rho_{\rm b}}{\rho_{\rm crit}(0)}, \ \ \ \
\Omega_{\Lambda}\equiv\frac{\Lambda}{3H_0^2}, \ \ \ \ \Omega_{\rm
K}\equiv\frac{-Kc^2}{H_0^2}
\end{equation}
where $\rho_{\rm b}$ is the mean baryonic matter density in the
universe at present time $t_0$ (redshift $z=0$), $\rho_{\rm
crit}(0)=3H_0^2/(8\pi G)=2.78\times 10^{11}h^2M_{\odot}$Mpc$^{-3}$
is the present critical mass density, and
$H_0=100h$kms$^{-1}$Mpc$^{-1}$ is the Hubble constant. We choose
$a(t_0)=1$. Since $\rmd\chi=c\rmd z/H(z)$, the proper distance from
the observer to an object at redshift $z$ is
$D^p(z)=c\int_0^z[(1+z)H(z)]^{-1}\rmd z$, where the Hubble parameter
at redshift $z$ is (known as Friedmann's equation)
\begin{equation}
H(z) \equiv \frac{\dot{a}}{a}=H_0\sqrt{\Omega_b(1+z)^3
+\Omega_K(1+z)^2+\Omega_\Lambda}.
\end{equation}
 The comoving distance from an
object at redshift $z_1$ to an object at redshift $z_2$ is
\begin{equation}
\chi(z_1,z_2)=\int_{z_1}^{z_2}\frac{c\rmd z}{H(z)},
\end{equation}
the corresponding angular diameter distance therefore is
\begin{equation}
D(z_1,z_2)=\frac{1}{1+z_2}f_K[\chi(z_1,z_2)].
\end{equation}

In TeVeS, the lensing equation has the same form as in general
relativity (GR),
 and for a spherically symmetric density profile \cite{zhao06a}
\begin{equation}
\beta=\theta-\frac{D_{LS}}{D_{S}}\alpha, \, \, \, \,
\alpha(b)=\int_0^{\infty}\frac{4b}{c^2r} \frac{\rmd\Phi(r)}{\rmd
r}\rmd l, \label{lens}
\end{equation}
where $\beta$, $\theta=b/D_{L}$ and $\alpha(\theta)$ are the source
position angle, image position angle and deflection angle,
respectively; $b$ is the impact parameter; $D_L$, $D_S$ and $D_{LS}$
are the angular diameter distances from the observer to the lens, to
the source and from the lens to the source, respectively;
$g(r)=\rmd\Phi(r)/\rmd r$ is the actual gravitational acceleration
[here $\Phi(r)$ is the spherical gravitational potential of the
lensing galaxy and $l$ is the light path]. It is well known that the
stellar component of an elliptical galaxy can be well modeled by a
Hernquist profile
\begin{equation}
\rho(r)=\frac{M_0r_h}{2\pi r(r+r_h)^3},
\end{equation}
with the mass interior to $r$ as
\begin{equation}
M(r)=\frac{r^2M_0}{(r+r_h)^2},
\end{equation}
where $M_0=\int_0^{\infty}4\pi r^2\rho(r)\rmd r$ is the total mass
and $r_h$ is the scale length. The corresponding Newtonian
acceleration is $g_N(r)=GM(r)/r^2=GM_0/(r+r_h)^2$. According to MOND
\cite{milgrom83,sanders02,sanders06}, the actual acceleration $g(r)$
is related to Newtonian acceleration by
\begin{equation}
g(r)\mu(g(r)/a_0)=g_N(r), \label{ac}
\end{equation}
where $\mu(x)$ is the interpolating function and has the properties
\begin{equation}
\mu(x)=\cases{x, & for $x\ll 1$ \cr 1, & for $x\gg 1$}
\end{equation}
 and $a_0=1.2\times 10^{-8}$cms$^{-2}$ is the critical acceleration below
 which gravitational law transits from Newtonian regime to MONDian regime.
 The concrete form of a $\mu(x)$ function should be determined by
observational data (e.g., the rotation curves of spiral galaxies)
and expected by a reasonable scalar field theory (e.g., TeVeS). The
``standard" function one usually takes is $\mu(x)=x/\sqrt{1+x^2}$,
which fits well to the rotation curves of most galaxies.
Unfortunately, if the MOND effect is produced by a scalar field
(such as TeVeS), the ``standard" $\mu(x)$ function turns out to be
multivalued \cite{zhao06b}. On the other hand, a ``simple" function
$\mu(x)=x/(1+x)$ suggested by Famaey \& Binney \cite{famaey05} fits
observational data better than the ``standard" function and is
consistent with a scalar field relativistic extension of MOND
\cite{zhao06b,sanders07}.

In order to explore a broad class of modified gravity models, Zhao
and Tian \cite{zhaotian06} proposed a parametrized modification
function
\begin{equation}
\frac{1}{\mu(g/a_0)}\equiv\frac{g}{g_N}=
\left[1+\left(\frac{a_0}{g_N}\right)^{kn}\right]^{\frac{1}{n}},
\label{gmu}
\end{equation}
in which, MOND gravity corresponds to $k=1/2$. Substituting equation
(\ref{ac}) into equation (\ref{gmu}) with $k=1/2$, we have
\begin{equation}
\mu(g/a_0)=\left[1+\left(\frac{a_0}{g\mu(g/a_0)}
\right)^{\frac{n}{2}}\right]^{-\frac{1}{n}}, \label{mond_gmu}
\end{equation}
which can be easily solved to obtain the usual form of the $\mu$
function for MOND \cite{zhaotian06}
\begin{equation}
\mu(x)=x\left[\frac{1}{2}+\sqrt{\frac{1}{4}+x^n}\right]^{-2/n}, \ \
\ \ x=\frac{g}{a_0}. \label{ap}
\end{equation}
It is easy to verify that the ``simple" and ``standard" $\mu$
function are approximated with high accuracy by equation (\ref{ap})
with $n=3/2$ and $n=3$, respectively \cite{zhaotian06}. The
requirement for a physical and monotonic $\mu$ function limits the
parameter $n$ to the range of $1.5\leq n\leq 2.0$. In this paper, we
consider three cases: $n=1.5$, 2.0 and 3.0.

Since the MONDian gravitational acceleration $g$ is explicitly
expressed in terms of the Newtonian acceleration $g_N$, it is very
convenient to use equation (\ref{gmu}) to calculate the deflection
angle
\begin{eqnarray}
\alpha(b)&=&\frac{4}{c^2}\int_0^{\infty}\frac{g(r)b}{r}\rmd l \nonumber \\
&=&\int_0^{\infty}\frac{4GM_0}{c^2}\frac{b}{r}\frac{1}{(r+r_{\rm
h})^2}[1+(\frac{a_0}{g_{\rm N}})^{n/2}]^{1/n}\rmd l
\end{eqnarray}
By using $r=b\sqrt{1+(l/b)^2}$ and $\theta=b/D_{\rm L}$, we have
\begin{equation}
\fl \alpha(\theta)=0^{''}.207h^{-1}\left(\frac{c/H_0}{D_{\rm
L}}\right)\frac{M}{\theta}\int_0^{\infty}\frac{[1+(a_0/g_{\rm
N})^{n/2}]^{1/n}}{\sqrt{1+x^2}[0.05r_{\rm h}(c/H_0)/(D_{\rm
L}\theta)+\sqrt{1+x^2}]}\rmd x, \label{alpha}
\end{equation}
where $M=M_0/M_{\star}$ and $M_\star=7.64\times
10^{10}h^{-2}M_{\odot}$ is the characteristic mass of galaxies
\cite{panters}, and
\begin{equation}
\frac{a_0}{g_{\rm N}}=2.38\left(\frac{D_{\rm
L}}{c/H_0}\right)^2\frac{\theta^2}{M}\left(\sqrt{1+x^2}+0.05\frac{c/H_0}{D_{\rm
L}}\frac{r_{\rm h}}{\theta}\right)^2. \label{ag}
\end{equation}
In equations (\ref{alpha}) and (\ref{ag}), the image position angle
$\theta$ and the scale length $r_{\rm h}$ are in units of arcsecond
( $^{''}$ ) and Kpc, respectively.

We need a relationship between the scale length $r_h$ and the mass
$M$, which could be determined by observational data. First, the
scale length is related to the effective (or half-light) radius
$R_e$ of a luminous galaxy by $r_h=R_e/1.8$ \cite{hernquist}.  It
has long been recognized that there exists a correlation between
$R_e$ and the mean surface brightness $\langle I_e\rangle$ interior
to $R_e$ \cite{djorgovski}: $R_e\propto\langle I\rangle_e^{-0.83\pm
0.08}$ . Since the luminosity interior to $R_e$ (half-light) is
$L_e=L/2=\pi\langle I\rangle_e R_e^2$, one immediately finds
$R_e\propto L^{1.26}$. Second, we need to know the mass-to-light
ratio $\Upsilon=M/L\propto L^{p}$ for elliptical galaxies. The
observed data gives $p=0.35$ \cite{van}; according to MOND, however,
we should find $p\approx 0$ \cite{sanders06}. In any case we have
\begin{equation}
L\propto M^{1/(1+p)}. \label{lm}
\end{equation}
Therefore, the scale length should be related to the stellar mass of
a galaxy by $r_h=AM^{1.26/(1+p)}$, and the coefficient $A$ should be
further determined by observational data. Without a well defined
sample at our disposal, we use the galaxy lenses which have an
observed effective radius $R_e$ (and thus $r_h$) in the CASTLES
survey \cite{munoz}, which are listed in table 2 of \cite{zhao06a}.
The fitted formulae for $r_{\rm h}$ are
\begin{equation}
r_{\rm h}=\cases{0.72\left(\frac{M}{M_{\star}}\right)^{1.26} \ \
{\rm Kpc}, & for p=0.0, \cr
1.24\left(\frac{M}{M_{\star}}\right)^{1.26/1.35} \ \ {\rm Kpc}, &
for p=0.35}. \label{rh}
\end{equation}
In later calculations, except indicated, we use the fitted formula
of $r_{\rm h}$ for $p=0$ as required by MOND.

\section{Galaxy stellar mass function}
In LCDM cosmology, mass function of virialized CDM halos can be
obtained in two independent ways. One is via the generalized
Press-Schechter (PS) theory, the other is via Schechter luminosity
function. In TeVeS, however, the PS-like theory does not exist.
Fortunately, the stellar mass function of galaxies is available in
the literature, including the one constrained by the most recent
data \cite{fontana06,panters}.

Before giving the galaxy stellar mass functions (GSMF) appeared in
the most recent literature, it is helpful to derive a GSMF directly
from the Schechter luminosity function and mass-to-light ratio. The
Schechter luminosity function is
\begin{equation}
\phi(L)=\phi^{\star}\left(\frac{L}{L_{\star}}
\right)^{\alpha}\exp\left(-\frac{L}{L_{\star}}\right)\frac{\rmd
L}{L_{\star}}. \label{schechter}
\end{equation}
For $L/L_{\star}=(M/M_{\star})^{1/(1+p)}$ implied by equation
(\ref{lm}), we have a GSMF
\begin{equation}
\phi(M)=\frac{\phi^{\star}}{1+p}\left(\frac{M}{M_{\star}}\right)^{\frac{\alpha+1}{1+p}-1}
\exp\left[-\left(\frac{M}{M_{\star}}\right)^{\frac{1}{1+p}}\right]\frac{\rmd
M}{M_{\star}}. \label{gsmf}
\end{equation}
While the average number density of galaxies $\phi_{\star}$, the
slope at low-mass end $\alpha$ and the slope of mass-to-light ratio
$p$ may be easily found from the published observational data or
assumptions, the characteristic stellar mass of galaxies $M_{\star}$
can be derived from
\begin{equation}
\rho_{\rm lum}=\Omega_{\rm lum}\rho_{\rm
crit}(0)=\int^{\infty}_0M\phi(M)\rmd M,
\end{equation}
where $\rho_{\rm lum}$ is the luminous baryonic matter density (note
that $\rho_{\rm lum}\ll\rho_{\rm b}$). The characteristic mass
$M_{\star}$ is
\begin{equation}
M_{\star}=\frac{\Omega_{\rm lum}\rho_{\rm
crit(0)}}{\phi_{\star}\Gamma(\alpha+p+2)}.
\end{equation}
For example, for $(\phi_{\star}, \alpha, \Omega_{\rm lum},
p)=(0.014h^3{\rm Mpc}^{-3}, -1.1, 0.003, 0.35)$ from \cite{lang},
$M_{\star}=6.56\times 10^{10}h^{-1}M_{\odot}$; for the same
parameters except that $p=0.0$ (MOND), $M_{\star}=5.56\times
10^{10}h^{-1}M_{\odot}$.

\begin{figure}[b]\center
\includegraphics[scale=0.4]{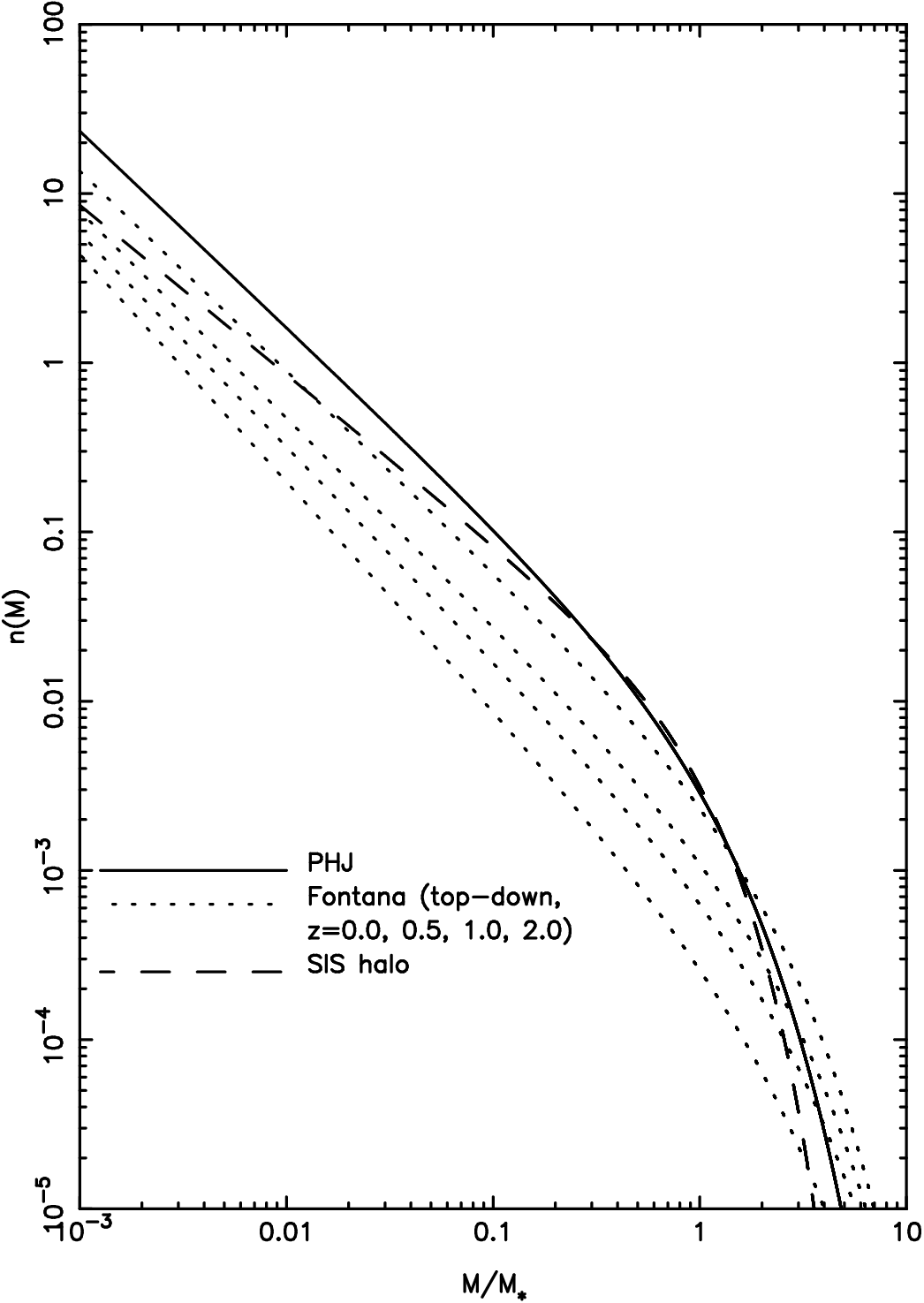}
\caption{Comoving number density for PHJ (solid), Fontana (dotted)
and SIS halos (dash). Since Fontana mass function depends on
redshift, four cases with redshift $z=0.0, 0.5, 1.0, 2.0$ are
displayed. For comparison, we normalize the three mass functions to
the same value of characteristic mass $M_\star=7.64\times
10^{10}h^{-2}M_{\odot}$.} \label{mf}
\end{figure}

Fortunately, the parameters in equation (\ref{gsmf}) have been
determined by recent observational data. By determining
non-parametrically the stellar mass functions of 96545 galaxies from
the Sloan Digital Sky Survey Data (SDSS) release one, Panter,
Heavens and Jimenez \cite{panters} (PHJ, hereafter) give the GSMF
\cite{chen06}
\begin{equation}
\phi(M)\rmd
M=\phi_\star\left(\frac{M}{M_\star}\right)^{\tilde{\alpha}}
\exp\left(-\frac{M}{M_\star}\right)\frac{dM}{M_\star}, \label{phj}
\end{equation}
where, we use $\phi(M)$ to denote the comoving number density 
of galaxies with mass between $M$ and $M+\rmd M$, and
\begin{equation}
\eqalign{\phi_\star=(7.8\pm 0.1)\times 10^{-3}h^3\mbox{Mpc}^{-3},
\cr \tilde{\alpha}=-1.159\pm 0.008, \cr M_\star=(7.64\pm 0.09)\times
10^{10}h^{-2}M_{\odot} .}
\end{equation}
Most recently, in order to study the assembly of massive galaxies in
the high redshift Universe, Fontana et al. \cite{fontana06}(Fontana,
hereafter) used the GOODS-MUSIC catalog to measure the evolution of
the GSMF and of the resulting stellar mass density up to redshift
$z=4$. The GSMF they obtained is
\begin{equation}
\phi(M,z)\rmd
M=\phi_{\star}(z)\left[\frac{M}{M_{\star}(z)}\right]^{\tilde{\alpha}(z)}
\exp\left[-\frac{M}{M_{\star}(z)}\right]\frac{\rmd M}{M_{\star}(z)},
\label{fontana}
\end{equation}
where
\begin{equation}
\eqalign{ \phi_{\star}(z)=n_0^{\star}(1+z)^{n_1^{\star}}, \
n_0^{\star}=0.0035, \  n_1^{\star}=-2.20\pm 0.18, \cr
\tilde{\alpha}(z)=\tilde{\alpha_0}+\tilde{\alpha_1}z, \
\tilde{\alpha_0}=-1.18, \tilde{\alpha_1}=-0.082\pm 0.033, \cr
M_{\star}(z)=10^{M_0^{\star}+M_1^{\star}z+M_2^{\star}z^2}h^{-2}M_{\odot},
\cr M_0^{\star}=11.16, M_1^{\star}=0.17\pm 0.05,
M_0^{\star}=-0.07\pm 0.01}
\end{equation}

It would be interesting to compare PHJ and Fontana GSMFs to the mass
function of galaxies in LCDM cosmology when the galactic halos are
modeled by SIS. The comoving number density of galactic halos with
velocity dispersion between $v$ and $v+\rmd v$
\cite{mitchell05,chen06} is
\begin{equation}
\phi(v)dv=\phi_{\star}\left(\frac{v}{v_{\star}}\right)^{\tilde{\alpha}}
\exp\left[-\left(\frac{v}{v_{\star}}\right)^{\tilde{\beta}}\right]
\tilde{\beta}\frac{v}{v_{\star}}, \label{sis}
\end{equation}
For comparison, we need to transform equation (\ref{sis}) from
velocity dispersion to halo mass $M$
\begin{equation}
M=4\pi\int_{0}^{r_{200}}\rho_{\rm
SIS}(r)r^2dr=\frac{800\pi}{3}r_{200}^3\rho_{\rm crit}(z),
\label{msis}
\end{equation}
where $r_{200}$ is the virial radius of a galactic halo within which
the average mass density is 200 times the critical density of the
Universe $\rho_{\rm crit}(z)$. Substituting the well known
expression $\rho_{\rm SIS}(r)=v^2/(2\pi Gr^2)$ into equation
(\ref{msis}), it is easy to find
\begin{equation}
\fl M(z)=6.58\times 10^5\left(\frac{v}{{\rm km}{\rm
s}^{-1}}\right)^3[\Omega_{\rm m}(1+z)^3+\Omega_{\rm
K}(1+z)^2+\Omega_{\Lambda}]^{-1/2}h^{-1}M_{\odot}, \label{mv}
\end{equation}
where $\Omega_{\rm m}$ is the matter density parameter (including
dark and baryonic components) \cite{li02}. Equation (\ref{mv}) means
that at any redshift $z$ we should have $M\propto v^3$, or for our
purpose, another form
\begin{equation}
\frac{M}{M_{\star}}=\left(\frac{v}{v_{\star}}\right)^3, \label{mv2}
\end{equation}
we thus have the galaxy mass function for SIS halos
\begin{equation}
\phi(M)=\frac{\phi_{\star}\tilde{\beta}}{3}
\left(\frac{M}{M_{\star}}\right)^{(\tilde{\alpha}-2)/3}
\exp\left[-\left(\frac{M}{M_{\star}}\right)^{\tilde{\beta}/3}\right]\frac{\rmd
M}{M_{\star}}. \label{sisms}
\end{equation}
We plot PHJ and Fontana GSMFs in figure 1 together with the galaxy
mass function for SIS halos (comoving number density). For SIS
halos, we use $(\phi_\star, \tilde{\alpha},
\tilde{\beta})=(0.0064h^3\mbox{Mpc}^{-3}, -1.0, 4.0)$ \cite{chae02}.
For comparison, we normalize the three mass functions to the same
value of characteristic mass $M_\star=7.64\times
10^{10}h^{-2}M_{\odot}$. Note that, for Fontana GSMF, the comoving
number density of galaxies decreases with increasing redshift, as
expected \cite{fontana06}.

\section{lensing probability}
Usually, lensing cross section defined in the lens plane with image
separations larger than $\Delta\theta$ is $\sigma(>\Delta\theta)=\pi
D_L^2\beta_{\rm cr}^2\Theta[\Delta\theta(M)-\Delta\theta]$, where
$\Theta(x)$ is the Heaviside step function and $\beta_{\rm cr}$ is
the caustic radius within which sources are multiply imaged. This is
true only when $\Delta\theta(M)$ is approximately constant within
$\beta_{\rm cr}$, and the effect of the flux density ratio $q_{\rm
r}$ between the outer two brighter and fainter images can be
ignored. Generally this is not true, readers are referred to
\cite{chen06} for details. We introduce a source position quantity
$\beta_{q_{\rm r}}$ determined by
\begin{equation}
\left(\frac{\theta(\beta)}{\beta}
\frac{d\theta(\beta)}{d\beta}\right)_{\theta>0}=q_{\rm r}
\left|\frac{\theta(\beta)}{\beta}
\frac{d\theta(\beta)}{d\beta}\right|_{\theta_0<\theta<\theta_{\rm
cr}}, \label{qr}
\end{equation}
where $\theta_0=\theta(0)<0$, the absolute value of which is the
Einstein radius, and $\theta_{\rm cr}$ is determined by
$d\beta/d\theta=0$ for $\theta<0$. Equation (\ref{qr}) means that
when $\beta_{q_{\rm r}}<\beta<\beta_{\rm cr}$, the flux density
ratio would be larger than $q_{\rm r}$, which is the upper limit of
a well defined sample. Therefore, the source position should be
within $\beta_{q_{\rm r}}$ according to the sample selection
criterion. For example, in the CLASS/JVAS sample, $q_{\rm r}\leq
10$.

The amplification bias should be considered in lensing probability
calculations.  For the source QSOs having a power-law flux
distribution with slope $\tilde{\gamma}$ ($=2.1$ in the CLASS/JVAS
survey), the amplification bias is
$B(\beta)=\tilde{\mu}^{\tilde{\gamma}-1}$ \cite{oguri02}, where, in
this paper,
\begin{equation}
\tilde{\mu}(\beta)=\left|\frac{\theta}{\beta}\frac{d\theta}{d\beta}
\right|_{\theta_0<\theta<\theta_{\rm
cr}}+\left(\frac{\theta(\beta)}{\beta}
\frac{d\theta(\beta)}{d\beta}\right)_{\theta>0}\label{mag}
\end{equation}
is the total magnification of the outer two brighter images. In our
previous work \cite{chen06}, however,  the amplification bias is
calculated based on the magnification of the second bright image of
the three images.

Therefore, the lensing cross section with image-separation larger
than $\Delta\theta$ and flux density ratio less than $q_{\rm r}$ and
combined with the amplification bias $B(\beta)$ is
\cite{schneider92,chenc,chen06}
\begin{eqnarray}
\fl \sigma(>\Delta\theta,<q_{\rm r})=&2\pi D_L^2\times \nonumber \\
&\cases{\int_0^{\beta_{q_{\rm r}}}\beta
\tilde{\mu}^{\tilde{\gamma}-1}(\beta)d\beta, & for
$\Delta\theta\leq\Delta\theta_0$, \cr \left(\int_0^{\beta_{q_{\rm
r}}}-\int_0^{\beta_{\Delta\theta}}\right)\beta
\tilde{\mu}^{\tilde{\gamma}-1}(\beta)d\beta, & for
$\Delta\theta_0<\Delta\theta\leq\Delta\theta_{q_{\rm r}}$, \cr 0, &
for $\Delta\theta>\Delta\theta_{q_{\rm r}}$,}  \label{cross}
\end{eqnarray}
where  $\beta_{\Delta\theta}$ is the source position at which a lens
produces the image separation $\Delta\theta$,
$\Delta\theta_0=\Delta\theta(0)$ is the separation of the two images
which are just on the Einstein ring, and
$\Delta\theta_{q_r}=\Delta\theta(\beta_{q_r})$ is the upper-limit of
the separation above which the flux ratio of the two images will be
greater than $q_{r}$.

The lensing probability with image separation larger than
$\Delta\theta$ and flux density ratio less than $q_{\rm r}$, in
TeVeS cosmology, for the source QSOs at mean redshift $z_s=1.27$
lensed by foreground elliptical stellar galaxies is
\cite{chenc,chend,chene,chen06}
\begin{equation}
\fl P(>\Delta\theta,<q_{\rm r})=
\int_0^{z_s}\frac{dD^{p}(z)}{dz}dz\int_0^{\infty}\phi(M, z)(1+z)^3
\sigma(>\Delta\theta, <q_{\rm r})dM, \label{prob}
\end{equation}
\begin{figure}[t]\center
\includegraphics[scale=0.4]{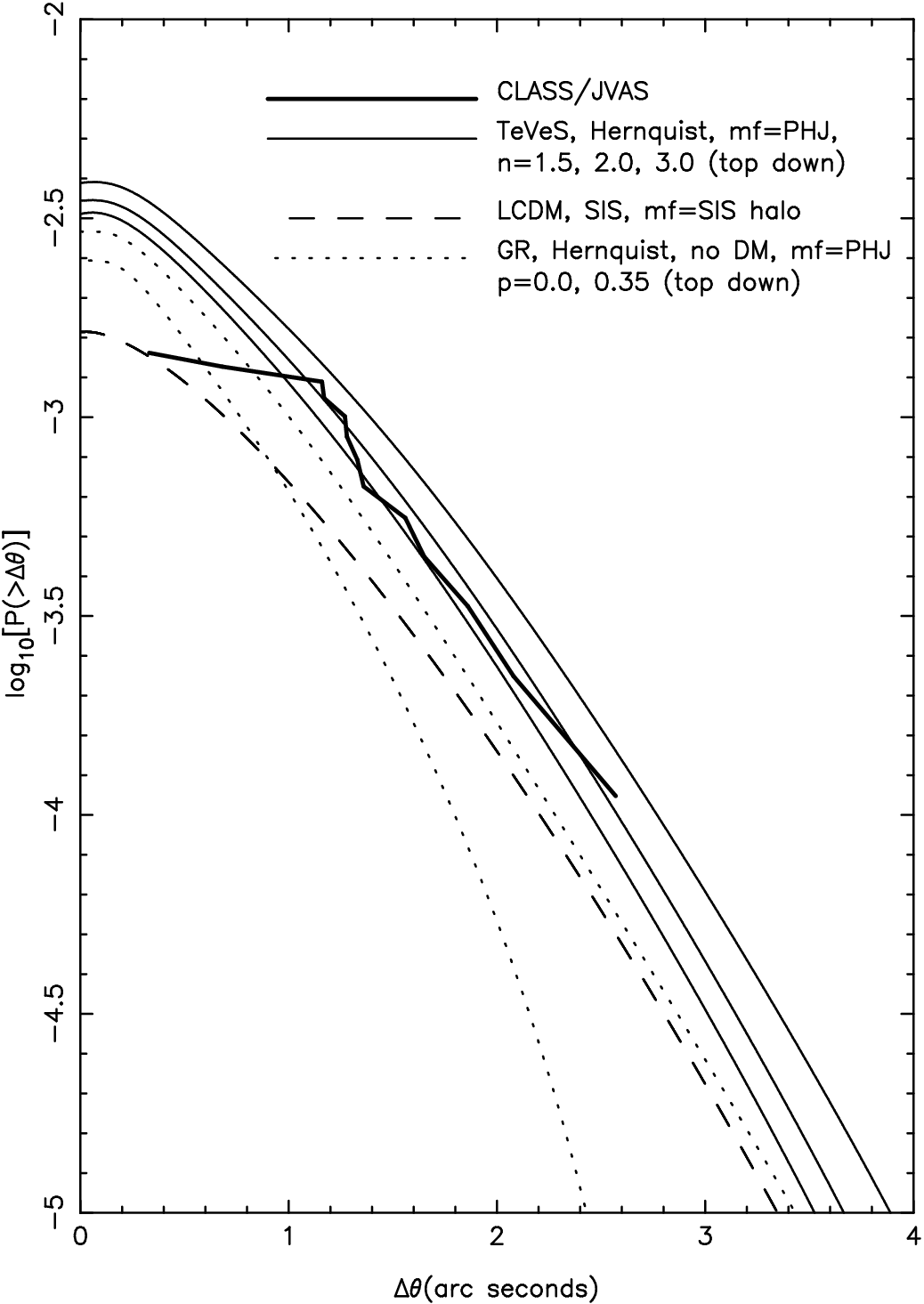}
\caption{Predicted lens probability with an image separation angle
$>\Delta\theta$ and the flux ratio $\le q_r=10$. For TeVeS (solid
line) and GR (no CDM and without modification of gravity, dotted
line), we assume an open cosmology with $\Omega_{\rm b}=0.04$ and
$\Omega_{\Lambda}=0.5$, model the lens as the Hernquist profile and
adopt PHJ GSMF (\ref{phj}); for standard LCDM (dashed line), we
assume a flat cosmology with $\Omega_{\rm m}=0.3$ and
$\Omega_{\Lambda}=0.7$, model the lens as the SIS and adopt the mass
function (\ref{sis}). For GR, we consider two different
mass-to-light ratio types and thus the expressions of $r_{\rm h}$,
see equation (\ref{rh}). For comparison, the survey results of
CLASS/JVAS (thick histogram) are also shown.} \label{prob_phj}
\end{figure}

\begin{figure}[t]\center
\includegraphics[scale=0.4]{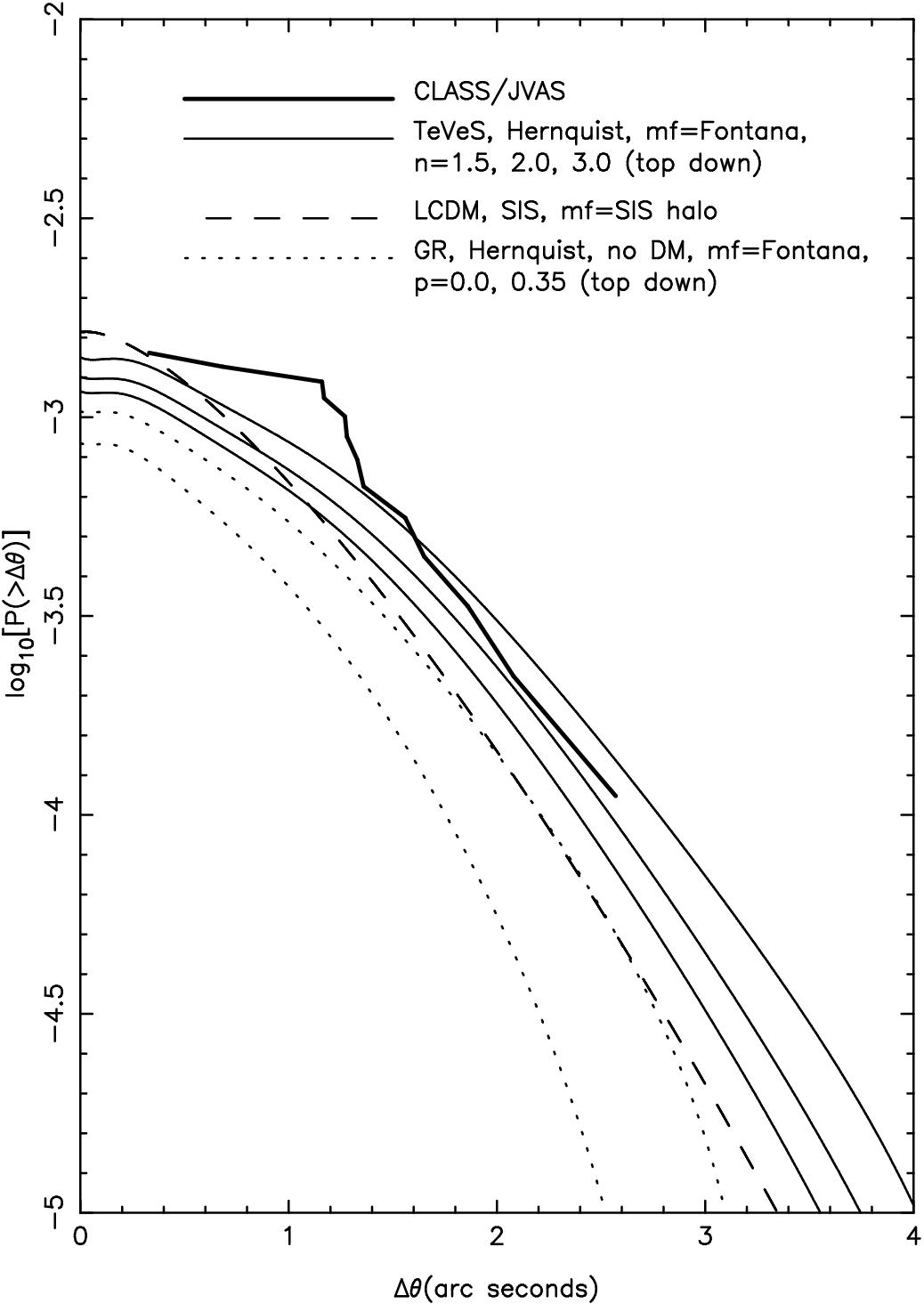}
\caption{same as Figure 2, except that GSMF=Fontana for TeVeS and
GR.} \label{prob_fontana}
\end{figure}

We plot in Figure \ref{prob_phj} and Figure \ref{prob_fontana} the
numerical results of the lensing probability according to equation
(\ref{prob}). In TeVeS (solid lines), we assume an open cosmology
with $\Omega_{\rm b}=0.04$ and $\Omega_{\Lambda}=0.5$,  as implied
by fitting to a high-z Type Ia supernova luminosity modulus
\cite{zhao06a}. The lensing galaxy is modeled by Hernquist profile
with length scale $r_{\rm h}=0.72(M/M_{\star})^{1.26}{\rm Kpc}$ for
constant mass-to-light ratio as required by MOND [see equation
(\ref{rh})]. The interpolating functions with three cases $n=1.5,
2.0$ and 3.0 are considered (top-down) according to equation
(\ref{mond_gmu}). In order to investigate the effects of MOND on
strong lensing, we also calculated the probabilities (dotted lines)
with no modification to gravitation theory (i.e., in GR) and without
dark matter (i.e., lensing galaxy is modeled by Hernquist profile).
In this case, two types of the fitted formulae for the length scale
$r_{\rm h}$ with $p=0$ and 0.35 (top-down) are adopted. In TeVeS and
GR (with no dark matter), we adopt the GSMF as the mass function
(mf), with mf=PHJ in Figure \ref{prob_phj} and mf=Fontana in Figure
\ref{prob_fontana}. As did in our previous work \cite{chen06},  We
recalculate the lensing probability with image separation larger
than $\Delta\theta$ and flux density ratio less than $q_r$, in flat
LCDM cosmology ($\Omega_m=0.3$ and $\Omega_\Lambda=0.7$), for the
source QSOs at mean redshift $z_s=1.27$ lensed by foreground SIS
modeled galaxy halos \cite{chae02,ma03,mitchell05}:
\begin{equation}
P_{\rm SIS}(>\Delta\theta,<q_{\rm
r})=\int_0^{z_s}dz\frac{dD^p(z)}{dz}\int_
{v_{\Delta\theta}}^{\infty}dv\bar{n}(v,z)\sigma_{\rm SIS}(v,z)B,
\end{equation}
where $\bar{n}(v,z)=\phi(v)(1+z)^3$, which is related to the
comoving number density $\phi(v)$ given by equation (\ref{sis}), is
the physical number density of galaxy halos at redshift $z$ with
velocity dispersion between $v$ and $v+dv$ \cite{mitchell05},
\begin{equation}
\sigma_{\rm
SIS}(v,z)=16\pi^3\left(\frac{v}{c}\right)^4\left(\frac{D_{\rm
LS}D_{\rm L}}{D_{\rm S}}\right)^2
\end{equation}
is the lensing cross section,
\begin{equation}
v_{\Delta\theta}=4.4\times
10^{-4}\left(\frac{c}{v_\star}\right)\sqrt{\frac{D_{\rm
S}\Delta\theta^{''}}{D_{\rm LS}}}
\end{equation}
is the minimum velocity for lenses to produce image separation
$\ge\Delta\theta^{''}$ and $B$ is the amplification bias. We adopt
$(\phi_\star,v_\star,\tilde{\alpha},\tilde{\beta})=(0.0064h^3\mbox{Mpc}^{-3},
198\mbox{kms}^{-1},-1.0,4.0)$ for early-type galaxies from
\cite{chae02}.  A subset of 8958 sources from
the combined JVAS/CLASS survey  form a well-defined statistical
sample containing 13 multiply imaged sources (lens systems)
suitable for analysis of the lens statistics
\cite{myers,browne03,patnaik92,king99}.  The observed lensing
probabilities can be easily calculated \cite{chenb,chenc,chene} by
$P_{\mathrm{obs}}(>\Delta\theta)=N(>\Delta\theta)/8958$, where
$N(>\Delta\theta)$ is the number of lenses with separation greater
than $\Delta\theta$ in 13 lenses. For comparison, the observational probability
$P_{\mathrm{obs}}(>\Delta\theta)$ for the
survey results of CLASS/JVAS is also shown (thick histogram). It would be
helpful for  us to figure out differences among models to summarize the 
values of the probabilities $P(>\Delta\theta=0.3'')$ in the Table 1.
\section{ Discussion and conclusions}

We have calculated the lensing probability with image separation
larger than a given value $\Delta\theta$ in an open, TeVeS
cosmology. The results are sensitive to the interpolating function
$\mu(x)$ and mass function $\phi(M,z)$. For a given GSMF (PHJ in
Figure \ref{prob_phj} and Fontana in Figure \ref{prob_fontana}), the
lensing probability decreases with increasing value of $n$ [given in
equation (\ref{ap})]. Obviously, for PHJ GSMF (Figure
\ref{prob_phj}), the lensing probabilities calculated in TeVeS
(solid lines for three cases of interpolating functions) are too
large at small lensing image separations compared with the results
of CLASS/JVAS. This unreasonable result is further confirmed, when
we note that, even the lensing probabilities in GR cosmology (with
no DM, dotted lines) are much larger than that in LCDM cosmology
(dashed line) at small image separations. Actually, however, this
result can be easly explained: at small mass-end (corresponding to
small image separation), the comoving number density for PHJ mf is
much larger than that for SIS halos (Figure \ref{mf}), which results
in the corresponding lensing probabilities according to equation
(\ref{prob}). This is why in our previous work \cite{chen06}, we
calculated the amplification bias based on the magnification of the
second bright image rather than the total magnification of the two
images considered. According to the resolution of CLASS/JVAS,
however, it is difficult to resolve the two images for small image
separations. Therefore, in this paper, we calculate the
amplification bias based on the total magnification of the outer two
brighter images, as usually done in the literature.

On the other hand, if we adopt another most recent mass function,
Fontana GSMF (Figure \ref{prob_fontana}), we find that the predicted
lensing probabilities in an open TeVeS cosmology with the ``simple''
interpolating function $\mu(x)=x/(1+x)$ [i.e., $n=3/2$ in equation
(\ref{ap})] match the observational data of CLASS/JVAS quite well.
Similarly, this is reasonable when we note that the comoving number
density of galaxies for PHJ GSMF is much higher than that for
Fontana GSMF at small mass-end (Figure 1). Clearly, the ``standard''
interpolating function $\mu(x)=x/\sqrt{1+x^2}$ [i.e., $n=3$ in
equation (\ref{ap})] is ruled out due to its too low lensing rates
at small image separations. Interestingly, this conclusion is in
agreement with the most recent result of Sanders and Noordermeer
\cite{sanders07}, who constrained the interpolating function with
the rotation curves of early-type disc galaxies.

\begin{table}[t]
\caption{The predicted values of the lensing probabilities $P(>\Delta\theta=0.3'')$ for all models and $P_{\rm  obs}(>\Delta\theta=0.3'')$ for CLASS/JVAS.}
\begin{tabular}{c|c|c|c|c|c|c|c} 
\br
             &\multicolumn{3}{c|}{TeVeS} & \multicolumn{2}{c|}{GR} & LCDM & CLASS/JVAS \\ 
\mr
             &n=1.5&n=2.0&n=3.0&p=0.0&p=0.35& &  \\ 
PHJ      & -2.460& -2.513& -2.552& -2.601& -2.691&-2.848&-2.838 \\  
Fontana&-2.873&-2.926& -2.965&-3.021&-3.112 &  &  \\ 
\br
\end{tabular}
\end{table}

In our calculations for deflection angle in TeVeS cosmology, we have
fixed the value of the critical acceleration $a_0$ and modeled the
lensing galaxies with the Hernquist profile, and the only free
choice is the interpolating function $\mu(x)$. We note that the PHJ 
GSMF includes all types of galaxies, 
whereas the mf for SIS halos includes only early-type galaxies, this can 
partly explain the relativey low abundance of SIS halos compared with PHJ GSMF in figure 1. 
On the other hand, Fontana GSMF, like PHJ GSMF,  also includes all types of galaxies, 
but its value is close to (when $z=0$)  or lower (for high $z$) than the mf for SIS halos.
Therefore, the major uncertainty
for lensing probability arises from the GSMF, which is
independent of any gravitational theory and should be determined by
observational data. Can we conclude from Figure 2 and Figure 3 that
Fontana GSMF is preferred and PHJ GSMF is ruled out? Recall that, in
LCDM cosmology, there are also uncertainties for the mass function
derived from the luminosity function, the equation (\ref{sis})
\cite{chae07,mitchell05,oguri07}. Actually, the parameters we adopted in equation
(\ref{sis}),
$(\phi_\star,v_\star,\tilde{\alpha},\tilde{\beta})=(0.0064h^3\mbox{Mpc}^{-3},
198\mbox{kms}^{-1},-1.0,4.0)$ \cite{chae02}, are selected so that
the predicted lensing probabilities $P_{\rm SIS}(>0.3'')$ can
exactly match the observed value $P_{\mathrm{obs}}(>0.3'')$, i.e.,
$P_{\rm
SIS}(>\Delta\theta=0.3'')=P_{\mathrm{obs}}(>\Delta\theta=0.3'')$, see Table 1. 
The most recent parameters derived from SDSS DR3 \cite{choi07} is 
$(\phi_\star,v_\star,\tilde{\alpha},\tilde{\beta})=(0.008h^3\mbox{Mpc}^{-3},
161\mbox{kms}^{-1},2.32,2.67)$, however, this will not affect our results.
One can see clearly from Figure 2 that, at larger image separations,
the predicted lensing probabilities (dashed line) are well bellow
the observed values, i.e., $P_{\rm
SIS}(>\Delta\theta>0.3'')<P_{\mathrm{obs}}(>\Delta\theta>0.3'')$. As
a matter of fact, in strong lensing statistics, one usually compares
the predicted cumulate lensing probability at the image separation
of $\Delta\theta=0.3''$, and regards the under-estimates at larger
image-separations to be unimportant. We note, however, that there is
an inflexion at $\Delta\theta=1.16''$ for
$P_{\mathrm{obs}}(>\Delta\theta)$ calculated from the well-defined
sample of CLASS/JVAS (thick histogram), and there are no physical
interpretations for the flat part of the line when
$\Delta\theta<1.16''$. Although the sample of CLASS/JVAS is
``well-defined", this should not include each detail such as the
inflexion, and other observations, like \cite{inada07}, would
provide more information at $\Delta\theta<1.16''$. So we can
reasonably guess that the correct observational data should avoid
the inflexion, and the trend of the rising probabilities with
smaller and smaller image-separations should continue inward at
$\Delta\theta<1.16''$. In this sense,  the predicted lensing
probabilities in TeVeS cosmology, with a PHJ GSMF and the
``standard" interpolating function, match the observational data
quite well as shown in Figure 2.

We also note that, in Figure 3, the lensing probabilities in GR
cosmology (with no DM, dotted lines) are much lower than the
observational data. This imply that, as an alternative to CDM, MOND
can sufficiently  account for the strong lensing observations.

\ack{I am grateful to the anonymous referee for good suggestions to improve the presentation of this paper. This work was supported by the National Natural Science Foundation of China under grant 10673012 and CAS under grant KJCX3-SYW-N2.}

\end{document}